\documentclass[12pt,preprint]{aastex}
\usepackage{natbib}
\usepackage{graphicx}
\usepackage{fixltx2e}
\usepackage{url}
\usepackage{placeins}
\usepackage{booktabs}
\usepackage{rotating,multirow} 

\def\teff{$T_{\rm eff}$}
\def\logg{$\log g$}
\def\feh{[Fe/H]}

\def\a0{$A_{\rm 0}$}

\def\ltsim{\ifmmode\stackrel{<}{_{\sim}}\else$\stackrel{<}{_{\sim}}$\fi}

\begin{document}

\title{Improving LSST Photometric Calibration with Gaia Data}

\author{
Tim Axelrod\altaffilmark{1} and Calder Miller\altaffilmark{2}
}

\altaffiltext{1}{Steward Observatory, University of Arizona}
\altaffiltext{2}{BASIS Tucson North}

\begin{abstract}
We consider the possibility that the Gaia mission can supply data which will improve the photometric calibration of LSST.   After outlining the LSST calibration process and the information that will be available from Gaia, we explore two options for using Gaia data. The first is to use Gaia G-band photometry of selected stars, in conjunction with knowledge of the stellar parameters $T_{eff}$, $\log g$, and $A_V$, and in some cases $Z$, to create photometric standards in the LSST u, g, r, i, z, and y bands.  We consider both main sequence (MS) stars and DA white dwarfs (WD).  The accuracies of the resulting standard magnitudes from MS stars are found to be insufficient to satisfy LSST requirements using Gaia data alone, but with the potential to do so when supplemented with ground based spectroscopy. The accuracies of the WD derived standards are generally adequate, but also require ground based spectroscopy.  The second option is combine the LSST bandpasses into a synthetic Gaia G band, which is a close approximation to the real Gaia G band.  This allows synthetic Gaia G photometry to be directly compared with actual Gaia G photometry at a level of accuracy which is useful for both verifying and improving LSST photometric calibration.
\end{abstract}


\section{Introduction}
To meet its science requirements, the Large Synoptic Survey Telescope (LSST) \citep{Ivezic2007} places stringent requirements on the accuracy of its photometric calibration.  The uniformity of photometric zeropoints across the sky is required to be 10 mmag in the grizy bands, and 20 mmag in u.   Additionally, color accuracy is required to be 5 mmag for colors that do not involve u, and 10 mmag for those that do.   A plan for achieving these requirements is presented in \cite{Jones2013}.  Achieving this plan, and verifying the accuracy of its results, relies on external data in the form of photometric standards. The goal of this paper is to explore the possibility that Gaia \citep{Perryman2001} can contribute to the external data needed for LSST photometric calibration.  

The paper is organized as follows.  Section \ref{sec:LSSTPhotoCal} presents an overview of the LSST photometric calibration process, and the role of external data.  Section \ref{sec:GaiaData}  is a brief summary of the Gaia data products that are relevant to our purpose.  Section \ref{sec:DirectGaiaStandards} presents the transformation of Gaia magnitudes into LSST magnitudes, and simulated results for both MS stars and WD.  Section \ref{sec:SyntheticG} presents the synthetic G band approach and simulated results for MS stars.  Section \ref{sec:Conclusion} summarizes our results, and implications for future work.

\section{LSST Photometric Calibration}
\label{sec:LSSTPhotoCal}
The overall strategy for LSST photometric calibration is based on \cite{StubbsTonry2006}.  They proposed 
{\it directly} measuring the system throughput as a function of wavelength, 
focal plane position, and time.  Further, the {\it
normalization} of the throughput in each observation (the gray-scale
zeropoint) and the {\it shape} of the throughput curve (the wavelength
dependent terms), are explicitly separated and measured with separate procedures
for both the telescope system response and the atmospheric transmission. 
Calibration systems based on this approach are already in use at PanSTARRS and DES. 

\noindent
The goal of the calibration process is to produce standard magnitudes, defined by
\begin{equation}
\label{eqn:stdMag}
m_b^{std} = -2.5\, log_{10} \left( {\int_0^{\infty} {F_\nu(\lambda) \,
  \phi_b^{std}(\lambda) \, d\lambda} } \over F_{AB}  \right)
\end{equation}
where $\phi_b^{std}(\lambda)$ is the normalized standard bandpass for band b, set a priori for the survey, $F_\nu(\lambda)$ is the object flux at the top of the atmosphere, and $F_{AB}$ = 3631 Jy. To calculate $m_b^{std}$ the calibration system must determine, for each observation, the actual realized bandpass from detected electrons to the top of the atmosphere, $\phi_b^{obs}(\lambda)$, and a gray zeropoint $Z_b^{obs}$.  With these quantities in hand, the standard magnitude is calculated by
\begin{equation}
\label{eqn:stdMagImpl}
m_b^{std} =  m_b^{inst} \, +2.5 \, log_{10} \,  \left( { \int_0^\infty {f_\nu(\lambda) \,
    \phi_b^{obs}(\lambda) \, d\lambda} \over \int_0^\infty {f_\nu(\lambda) \,
    \phi_b^{std}(\lambda) \, d\lambda}} \right) + Z_b^{obs}
\end{equation}
\noindent
where $m_b^{inst}$ is the instrumental magnitude (log of counts) and $f_\nu(\lambda)$ is the {\it shape} of the source spectrum, which we will refer to as the SED.

\noindent
Several hardware systems are required to implement the approach, which we briefly describe:

\begin{itemize}
\item{A dome screen projector designed to provide uniform
    ($\sim10\%$ variation) illumination across the field of view, while
    minimizing stray light. This projector system will have the
    capability to not only illuminate the screen with broadband white
    light, but also narrow-band light to measure the system response
    at individual wavelengths. The narrow-band light will be generated
    by a tunable laser, capable of producing light from $300-1100$~nm
    and tunable in 1~nm increments. The brightness of the screen is measured 
    with a NIST-calibrated photodiode, so that the relative intensity at different 
    wavelengths can be precisely determined. } 
\item{A 1.2-m auxiliary telescope with an $R \approx 400$ spectrograph,
    located adjacent to the LSST itself. This auxiliary telescope will
    obtain spectra of a chosen set of atmospheric probe stars across the sky to
    determine an atmospheric absorption model.} 
\item{Water vapor monitoring system, consisting of a GPS system and a 
    microwave radiometer copointed with the LSST telescope, and monitoring 
    a similar field of view.  This supplements the auxiliary telescope spectra, which are
    unable to track the sometimes rapid variations of water vapor in time and space.} 
\end{itemize}

LSST uses a form of self calibration, as pioneered by \cite{Padmanabhan2008}, to set the zero point for each exposure.  The inputs to the self calibration process are the SED-corrected magnitudes of the calibration stars in band $b$, $m_b^{corr}$, for the entire survey.  These are calculated from the instrumental magnitudes $m_b^{inst}$, the object SEDs $f_\nu(\lambda,t)$, the standard bandpass for the photometric system, $\phi_b^{std}(\lambda)$, and the system bandpass at the moment of observation, $\phi_b^{obs}(\lambda,t)$ as
   
\begin{equation}
m_b^{corr} = m_b^{inst} -2.5 \, log_{10} \,  \left( { \int_0^\infty {f_\nu(\lambda) \,
    \phi_b^{obs}(\lambda) \, d\lambda} \over \int_0^\infty {f_\nu(\lambda) \,
    \phi_b^{std}(\lambda) \, d\lambda}} \right)
\end{equation}

Self calibration utilizes the facts that the survey observes each calibration star many times over the course of the survey, and that the survey fields overlap, to construct a least squares system for the zero points of each observation. The resulting set of zero points for each band has an arbitrary offset, which makes it impossible to determine object colors without additional information.  This additional information can come from a variety of sources:

\begin{itemize}
\item Photometric standards in individual LSST bands.
\item Photometric standards in a bandpass which can be synthesized from the LSST bands.
\item Locus of a well understood stellar population in color-color space.  
\end{itemize}

The need for external information goes beyond simply determining the band zero point offsets, which, in principle, could be obtained from only a single standard.  In practice, results from self calibration will contain some level of spatially dependent error, which tends to be greatest for the low-order spatial modes.  Assessing the magnitude of this error, and correcting it if needed, can only be performed if one or more of the above sources of information is available as a function of position on the sky.  This is, of course, the more demanding application, and it is the one we focus on in this paper.

As we show below, Gaia data can contribute both to band zero point determination and to verification/correction of self calibration errors through the first two of the approaches listed above.  We do not discuss the third approach here, but refer the reader to \cite{High2009} as an example of its use.

\section{Gaia Photometry and Stellar Parameters}
\label{sec:GaiaData} 

Gaia's principal photometric data product \citep{Perryman2001} is calibrated magnitudes through a broadband G filter.  Additionally, it produces low resolution red (RP) and blue (BP) spectra, and higher resolution (RVS) spectra within a narrow band intended mostly to supply radial velocities.  It is tempting to use the RP and BP spectra directly to estimate the stellar SED $F_{\nu}(\lambda)$ and then use that to calculate the color terms needed to transform from the Gaia G band to the 6 LSST bands.  The biggest problem with this approach is that RP and BP spectra, due to their low resolution and relatively broad PSF, cannot be directly inverted to give the incident stellar SED with sufficient accuracy. Preliminary simulations showed that systematic errors of order 0.1 mag are incurred, which rules out the direct approach for our purposes. We are better off taking a different approach, and instead using Apsis \citep{Bailer-Jones2013}, Gaia's astrophysical parameters inference system.  We realized that Apsis, which of course
utilizes BP/RP as a major input, would lead to smaller errors for our case. There
are two reasons for this.  First, Apsis takes account of data beyond
BP/RP, in particular the parallax to determine absolute luminosity.
Second, Apsis has built in the knowledge that these are not arbitrary
spectra, but the spectra of stars. This prior knowledge significantly
reduces the errors in the color terms $C_{bG}$ (defined in Section \ref{sec:Transforming}) from our initial naive concept.

Apsis utilizes all available Gaia data for each star to infer the stellar parameters $T_{eff}$, $\log g$, $[Fe/H]$, and $A_0$. $A_0$ is a stellar extinction parameter used by Apsis, which is within a few percent of the more commonly used extinction at 550 nm, $A_V$, for values of $R_V = A_V/E(B-V)$ near 3.  For our purposes the difference between $A_0$ and $A_V$ is inconsequential, and in the remainder of the paper we use the more common $A_V$.  For sufficiently bright stars ($G \leq 15$), this includes spectra from the RVS instrument, which substantially improves the accuracy of the resulting stellar parameters.  The usable range of magnitudes for LSST standards is roughly $18 < r < 16$, with the bright end set by detector saturation, and the faint end set by signal-to-noise for a 15 sec exposure.  Therefore, we must make use of stellar parameters determined without RVS.  Table \ref{table:Apsis_accuracy} shows the predicted accuracy of these parameters.

\section{Creating Direct LSST Band Photometric Standards from Gaia Data}
\label{sec:DirectGaiaStandards}
As mentioned in Section \ref{sec:LSSTPhotoCal}, the errors remaining after self calibration can be assessed and/or eliminated if we have an accurate set of photometric standards in the six LSST bands with a suitable distribution across the sky.  Work is in progress \citep{Saha2013} to obtain a small set of suitable white dwarfs observed with HST whose photometry can be accurately transformed to the LSST bands, but their number and spatial distribution may not be adequate to the task addressed here. The minimum number of standards required depends on the spatial power spectrum of the self calibration errors, and is difficult to determine in the absence of actual LSST data.  The low order spatial error modes ($L<3$) from self-calibration can be adequately measured over the roughly $2 \pi$ steradians observed by LSST with roughly ten standards.   Clearly, it would be desirable to have more, if only to verify that the power spectrum is negligible at higher $L$ values.  As an upper limit, 2000 standards would provide one standard in each LSST field of roughly 10 square degrees.  Every LSST observation would then include a standard, and spatial error modes to $L \le 40$ would be tightly constrained.  The number of standards employed in practice will be guided by the error characteristics of the self calibration process, and possibly constrained by the availability of ground-based spectroscopy resources (see Section \ref{sec:UsingWD})

Each observation of a standard constrains the self calibration equations by a direct measurement of the zero point at that place and time:

\begin{equation}
Z_b(t) =  -2.5 \, log_{10} \int_0^\infty {F_\nu^{std}(\lambda) \,
    \phi_b^{obs}(\lambda,t) \, d\lambda} - m_b^{inst}
\end{equation}

In this paper we are concerned with the uncertainties in determining $Z_b(t)$ that result solely from uncertainties in $F_\nu^{std}(\lambda)$.  In evaluating these, we make two assumptions:
\begin{itemize}
\item $\phi_b^{obs}$ is known exactly.  This is of course not the case, since it depends on a variety of measurements with finite uncertainty, but this allows us to isolate the uncertainty which is due solely to the precision of the standard.
\item $\phi_b^{obs}$ is sufficiently close to $\phi_b^{std}$ that the difference can be ignored in this context.  This is not obviously the case, since $\phi_b^{std}$ is in principle arbitrary, but in the case of LSST we will pick $\phi_b^{std}$ to be statistically as close as possible to $\phi_b^{obs}$.
\end{itemize}

We therefore approximate the calibration errors that result from an imperfectly characterized standard as
\begin{equation}
\delta Z_b(t) =  -2.5 \, log_{10} \int_0^\infty { \delta F_\nu^{std}(\lambda) \,
    \phi_b^{std}(\lambda,t) \, d\lambda}
\end{equation}

where $\delta F_\nu^{std}(\lambda)$ is the difference between the actual and assumed SED of the standard.

\subsection{Transforming Gaia Magnitudes into LSST Magnitudes}
\label{sec:Transforming}
\begin{figure}
\plotone{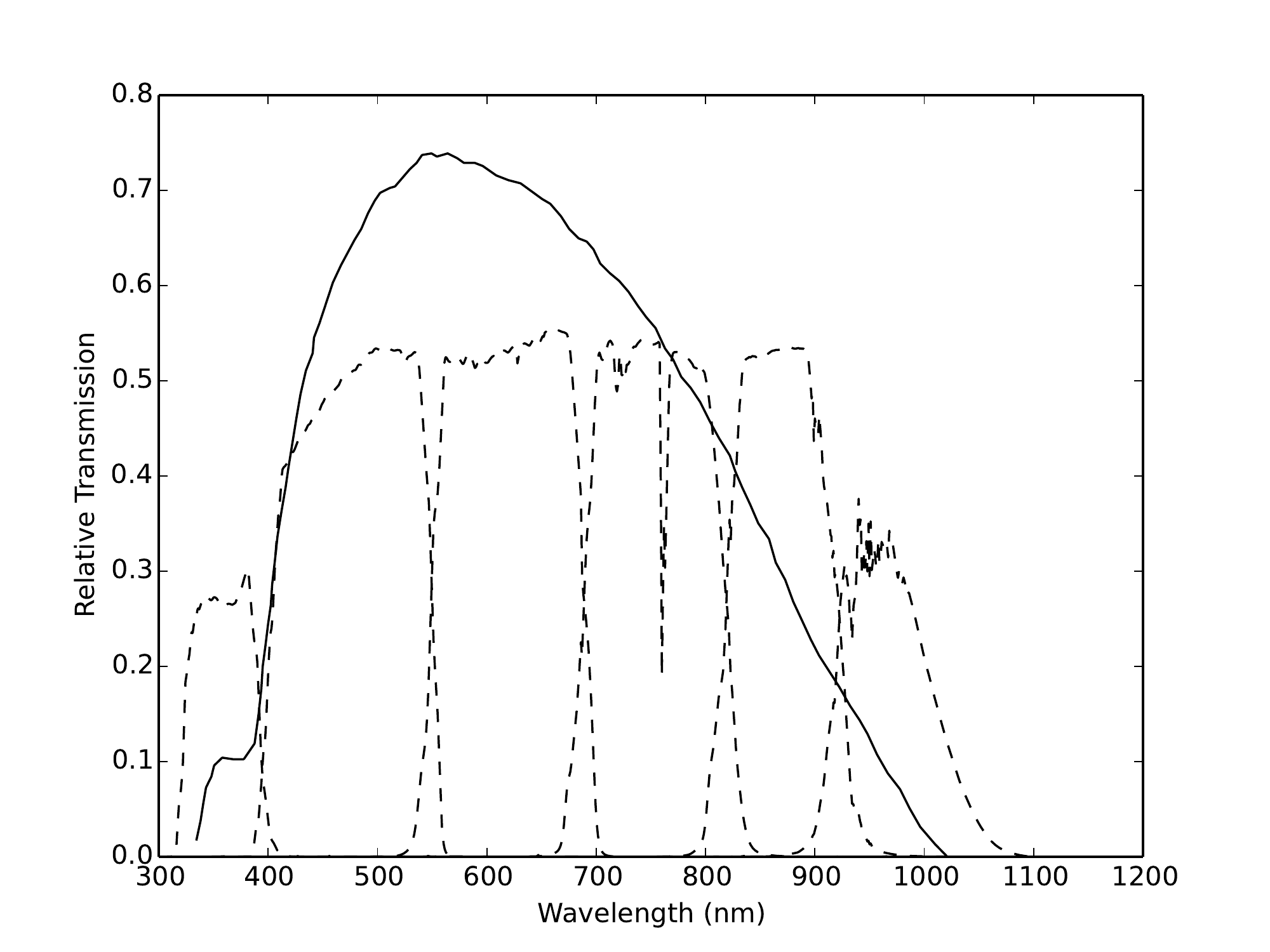}
\caption{Bandpasses for Gaia-G band (solid curve), and LSST u, g, r, i, z, and y bands (dotted curves, from left to right).  The LSST bands include a reference atmospheric transmission, as well as nominal transmissions for all optical components, and the detector quantum efficiency.}
\label{fig:bandpasses}
\end{figure}

To make use of Gaia data for LSST photometric standards, we must transform Gaia G magnitudes into LSST magnitudes in each of the six LSST filter bands.  As Figure \ref{fig:bandpasses} shows,  the Gaia G band is very broad compared with the LSST bands, and has an overlap with each of them.  This suggests that an accurate transformation might be possible given an achievable level of knowledge of the stellar SED.  The transformation process is as follows:
\begin{equation}
m_{b} = m_{G} + C_{bG}
\end{equation}
where $b$ denotes one of the six LSST bands, and $G$ the Gaia G band.
\begin{equation}
C_{bG} \equiv m_{LSST_b} - m_{Gaia_G} = -2.5 \, log_{10} \,  \left( { \int_0^\infty {F_\nu(\lambda) \,
    \phi_b(\lambda) \, d\lambda} \over \int_0^\infty {F_\nu(\lambda) \,
    \phi_G(\lambda) \, d\lambda}} \right)
\label{eqn:GaiaLsstColor}
\end{equation}
We assume that the stellar flux, $F_\nu(\lambda)$, can be adequately parameterized by the Apsis stellar parameters $T_{eff}, \log g, Z, A_V$, which in turn means that the color $C_{bG}$ can also be so parameterized:
\begin{equation}
C_{bG} = C_{bG}(T_{eff}, logg, Z, A_V)
\end{equation}

Our main concern in this paper is with the uncertainty in the resulting $LSST_b$ magnitudes when the stellar parameters are determined by Apsis.  The uncertainty is given by 

\begin{equation}
\sigma^{2}_{b} = \sigma^{2}_{G} + \sigma^{2}_{C_{bG}}
\end{equation}

\begin{figure}
\plotone{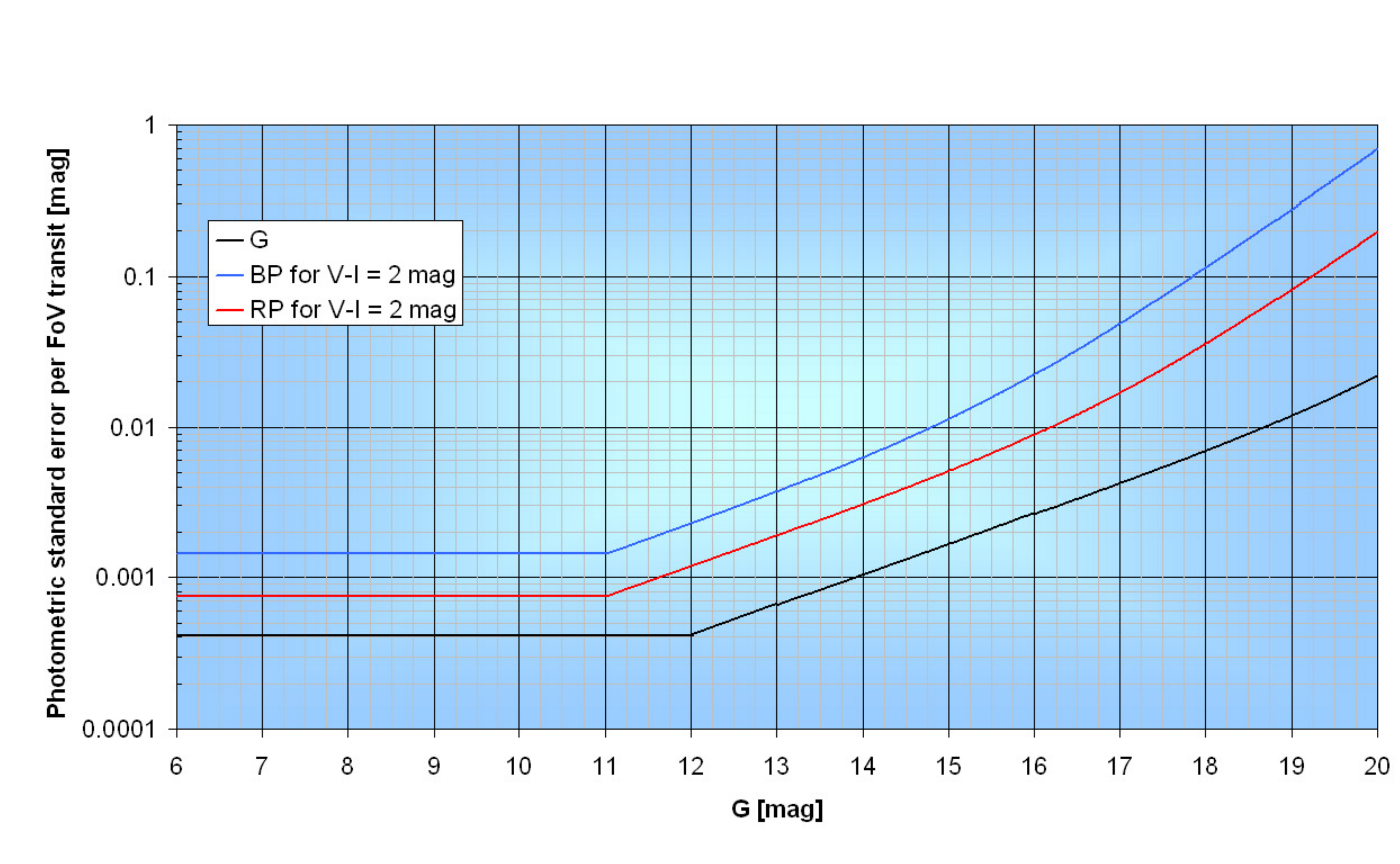}
\caption{Expected Gaia G band photometric accuracy for V-I = 2 mag, from \cite{deBruijne2012}.  Image courtesy ESA}
\label{fig:GaiaPhotometricErrors}
\end{figure}

The expected uncertainty in Gaia photometry, $\sigma_{G}$, is presented in \cite{deBruijne2012}, and reproduced in Figure \ref{fig:GaiaPhotometricErrors}.  Note that the uncertainties in the figure are for single visit photometry.   The expected uncertainty in the end of mission photometry is lower by roughly $1/\sqrt{70}$, yielding uncertainties of roughly 0.4 mmag at G=16 and 0.8 mmag at G=18.  It is clear that $\sigma_{G}$ makes a negligible contribution to the total uncertainty of 5 mmag required for LSST, as discussed in Section \ref{sec:LSSTPhotoCal}. Therefore we require the uncertainty $\sigma_{C_{bG}}$ to be 5 mmag or less. The remainder of the paper is largely occupied with the assessment of $\sigma_{C_{bG}}$.

\FloatBarrier
\subsection{Using Gaia Data on MS Stars to Calibrate LSST}
\label{sec:MS_Results}

Given the enormous number of stars in both the LSST and Gaia catalogs in the relevant magnitude range, we can be highly selective in the stellar types we select as standards.  The most useful stellar type for our purpose is not immediately clear, since it depends on the details both of the performance of Gaia Apsis, and on the shape of the LSST filter bandpasses.  We first examine the suitability of main sequence (MS) stars.

In order to estimate the precision to which LSST magnitudes can be found using Gaia's Apsis, we simulate the observed magnitudes of the stars in the Phoenix spectral library \citep{Brott2005} for the Gaia G band and the LSST u, g, r, i, z and y bands.  The Phoenix spectra do not include interstellar reddening due to dust, so we redden them with the prescription of \cite{Cardelli1989} using $R_V=3.1$.  

We then used the published uncertainties in the stellar parameters determined by Apsis (in $T_{eff}$, $\log g$, $z$ and $A_V$) to find the standard deviation of the colors $C_{bG}$ for objects whose parameters are within the Apsis uncertainties.  In this fashion, the standard deviation and other statistics concerning the predicted precision of Gaia measurements of stars can be found. We neglect the covariance between the various stellar parameters, which surely results in an overestimate of the dispersion in the resulting colors.  This assumption is necessary due to our lack of knowledge of the covariance terms, which have not yet been published.  The overestimate is likely in the range of 2 to 4. 

It is necessary to interpolate between the Phoenix data points because the Phoenix spectra used as input are widely spaced in stellar parameter space, so that typically only a few spectra are within the range of uncertainty of the Apsis stellar parameters.  The interpolation program linearly interpolates in $T_{eff}, \log g, Z$ and $A_V$ between the magnitudes calculated from the available spectra. The program selects the interpolated points drawn from a 4-dimensional normal distribution with variances determined by the expected Apsis errors, and then calculates the LSST magnitudes for each point.  It then finds the standard deviation of the $C_{bG}$ colors of the indistinguishable points, in order to find the predicted error from Gaia data.

\begin{table}
\begin{center}
\caption{Accuracy of astrophysical parameter estimation (internal RMS errors) with the Apsis algorithm Aeneas using
BP/RP spectra and parallaxes, but without RVS data \citep{Bailer-Jones2013}.
\vspace{10pt}
\label{table:Apsis_accuracy}
}
\begin{tabular}{crrccc}
\toprule
& $G$ & \teff\ & \a0\ & \logg\ & \feh\ \\
& mag & K & mag & dex & dex \\
\midrule
\multirow{3}{*}{\begin{sideways}\begin{footnotesize}A stars\end{footnotesize}\end{sideways}}
 & 9 & 340 & 0.08 & 0.43 & 0.86 \\
 & 15 & 260 & 0.06 & 0.38 & 0.93 \\
 & 19 & 400 & 0.15 & 0.51 & 0.74 \\
\midrule
\multirow{3}{*}{\begin{sideways}\begin{footnotesize}F stars\end{footnotesize}\end{sideways}}
 & 9 & 150 & 0.06 & 0.36 & 0.36 \\
 & 15 & 170 & 0.07 & 0.38 & 0.33 \\
 & 19 & 630 & 0.35 & 0.37 & 0.60 \\
\midrule
\multirow{3}{*}{\begin{sideways}\begin{footnotesize}G stars\end{footnotesize}\end{sideways}}
 & 9 & 140 & 0.07 & 0.31 & 0.14 \\
 & 15 & 140 & 0.07 & 0.22 & 0.16 \\
 & 19 & 450 & 0.33 & 0.45 & 0.65 \\
\midrule
\multirow{3}{*}{\begin{sideways}\begin{footnotesize}K stars\end{footnotesize}\end{sideways}}
 & 9 & 100 & 0.09 & 0.26 & 0.19 \\
 & 15 & 90 & 0.08 & 0.26 & 0.21 \\
 & 19 & 230 & 0.23 & 0.36 & 0.48 \\
\midrule
\multirow{3}{*}{\begin{sideways}\begin{footnotesize}M stars\end{footnotesize}\end{sideways}}
 & 9 & 60 & 0.13 & 0.15 & 0.21 \\
 & 15 & 70 & 0.14 & 0.29 & 0.25 \\
 & 19 & 90 & 0.13 & 0.17 & 0.29 \\
\bottomrule
\end{tabular}
\end{center}

\end{table}

\subsection{Results for MS Stars}
\begin{figure}[htbp]
\centering
\includegraphics[width=3in]{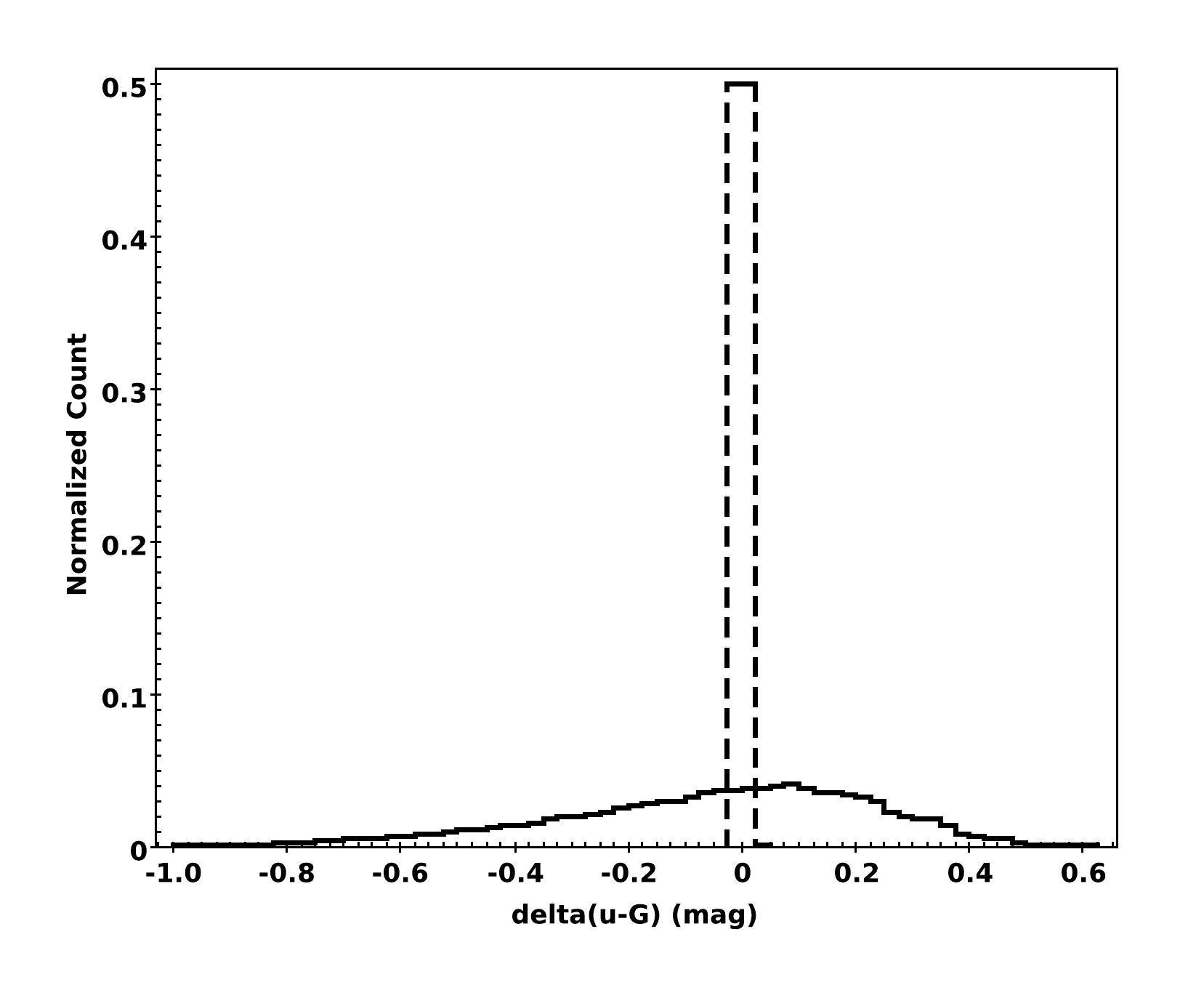}\includegraphics[width=3in]{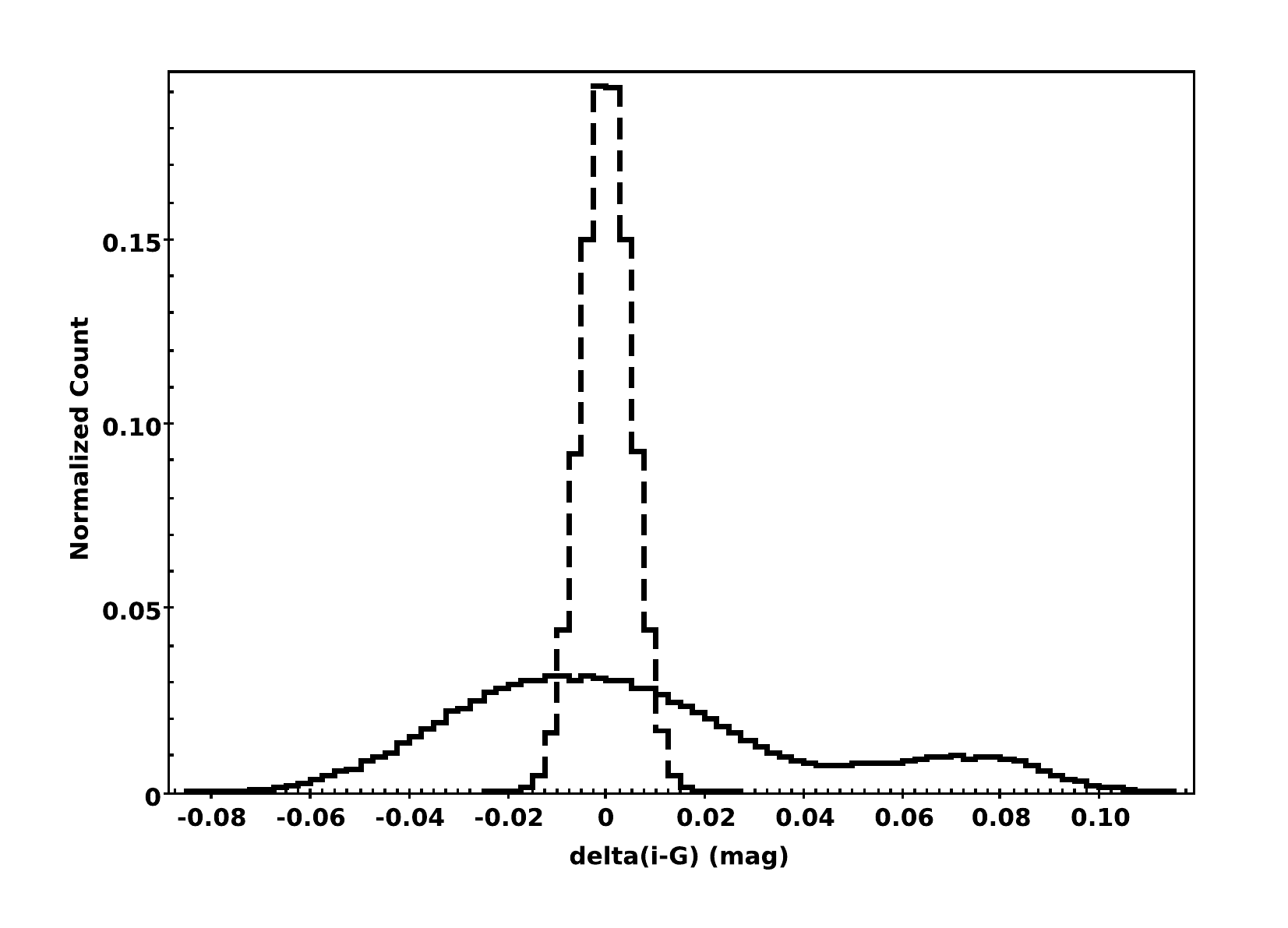} \\
\includegraphics[width=3in]{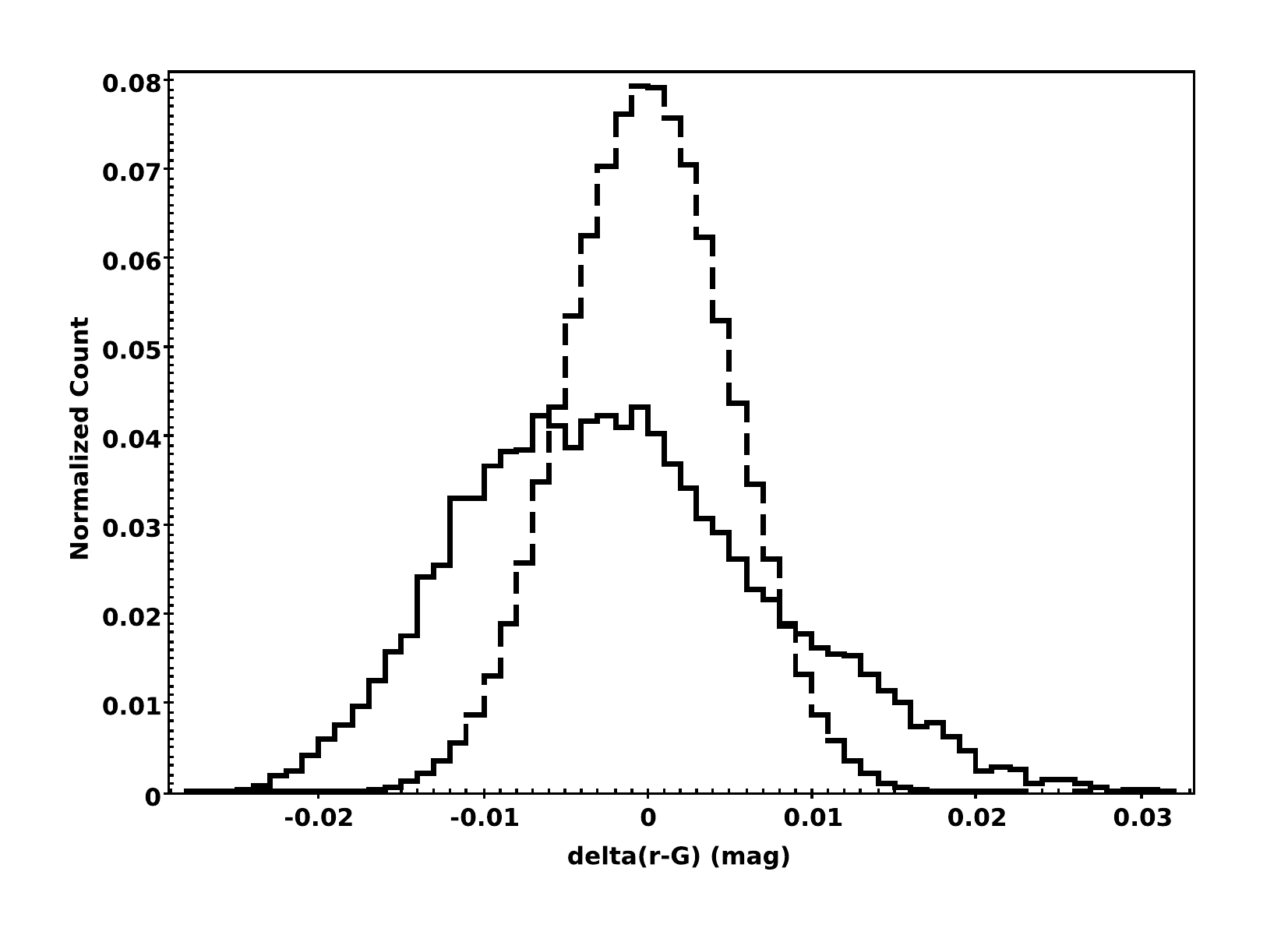}\includegraphics[width=3in]{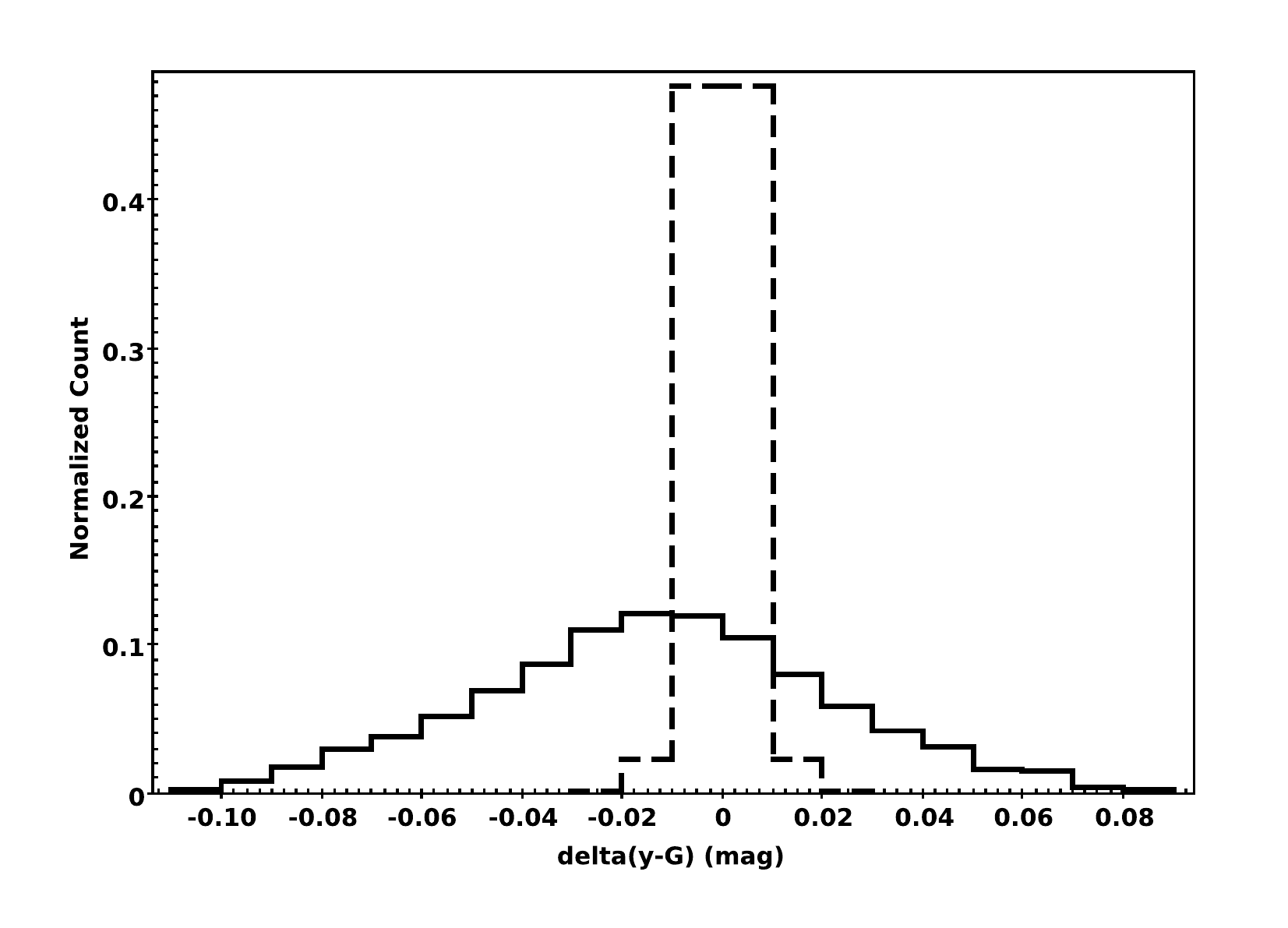}
\caption{ {Sampling of histograms of $C_{bG}$ colors for stars within the Apsis uncertainty window around a point in stellar parameter space.  Top row: $T_{eff}$=4437, $\log g$=4.73,$z$=-0.5, $A_V$=0.5.  Bottom row: $T_{eff}$=8472, $\log g$=4.41, $z$=-0.5, $A_V$=0.5.  A fixed width normal distribution with $\sigma$=0.005 mag is superposed on all.} 
\label{fig:MShistos} }
\end{figure}

Initial exploration of the stellar parameter space for main sequence stars showed that stellar temperature is the main variable which affects the accuracy of $C_{bG}$.  We present results from two points chosen from the SDSS DR8 stellar locus in \teff - \logg \hspace{1ex}space, a low temperature point at $T=4437K$, $\log g=4.73$, and a high temperature point at $T=8472K, \log g=4.41$.  The adopted Apsis uncertainties were $\sigma_T=160, \sigma_Z = .345, \sigma_{\log g} = .31, \sigma_{A_V} = .155$ for the low temperature point, and $\sigma_T=330, \sigma_Z = .835, \sigma_{\log g} = .445, \sigma_{A_V} = .105$ for the high temperature point. The results are best summarized as histograms of $C_{bG}$ for all six LSST bands.  The histograms are typically close to Gaussian, and are characterized by their standard deviations, $\sigma$, listed for both cases in Table 2. The increased sensitivity of the colors at low temperatures is evident.  A sampling of the histograms from which $\sigma$ is calculated are shown in Figure \ref{fig:MShistos}. A few of the low temperature histograms have a double peaked shape, which is an artifact that arises because the relatively coarse grid of Phoenix spectra does not smoothly resolve the rapid changes in the spectra at low temperature.

\begin{figure}[htbp]
\centering
\plottwo{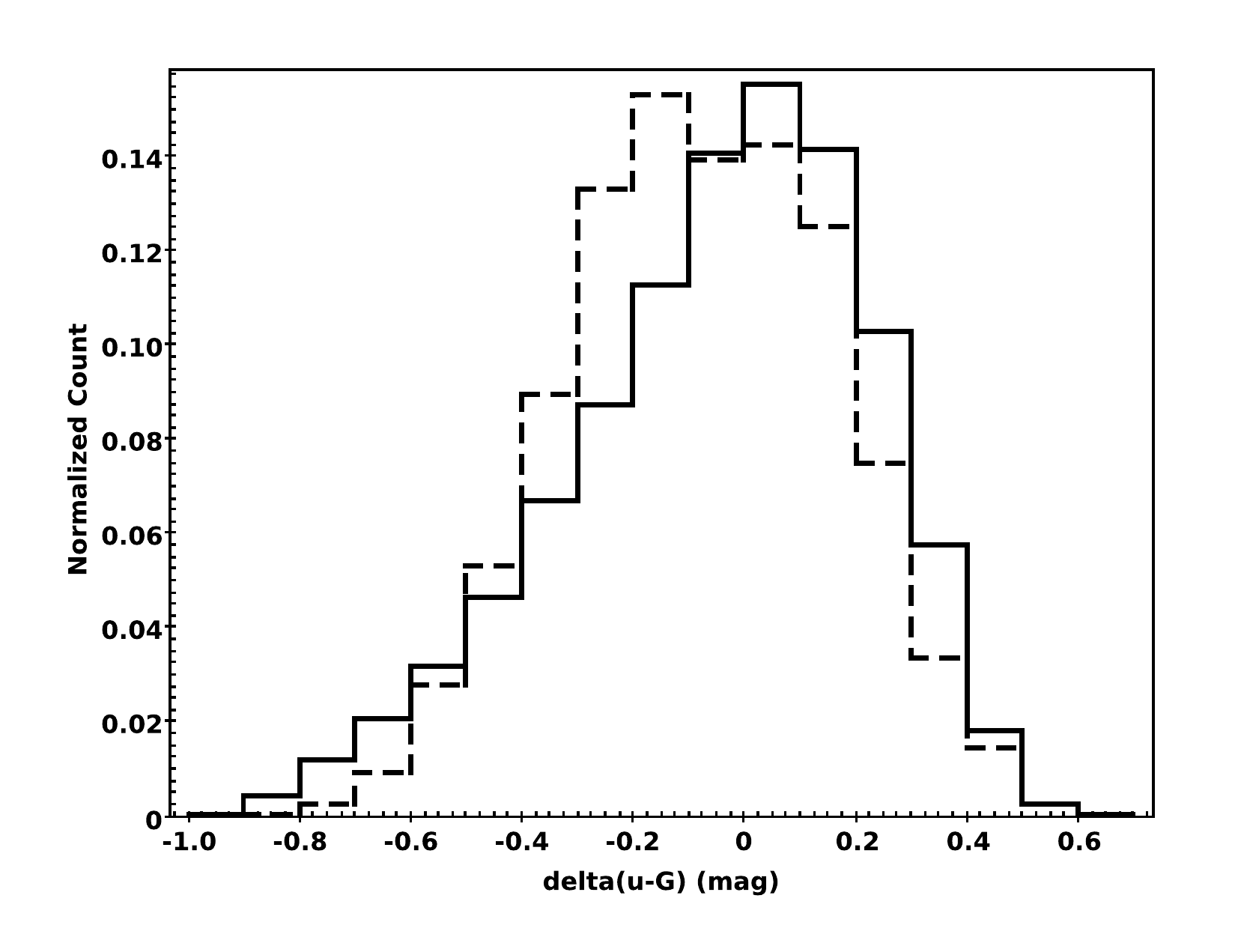}{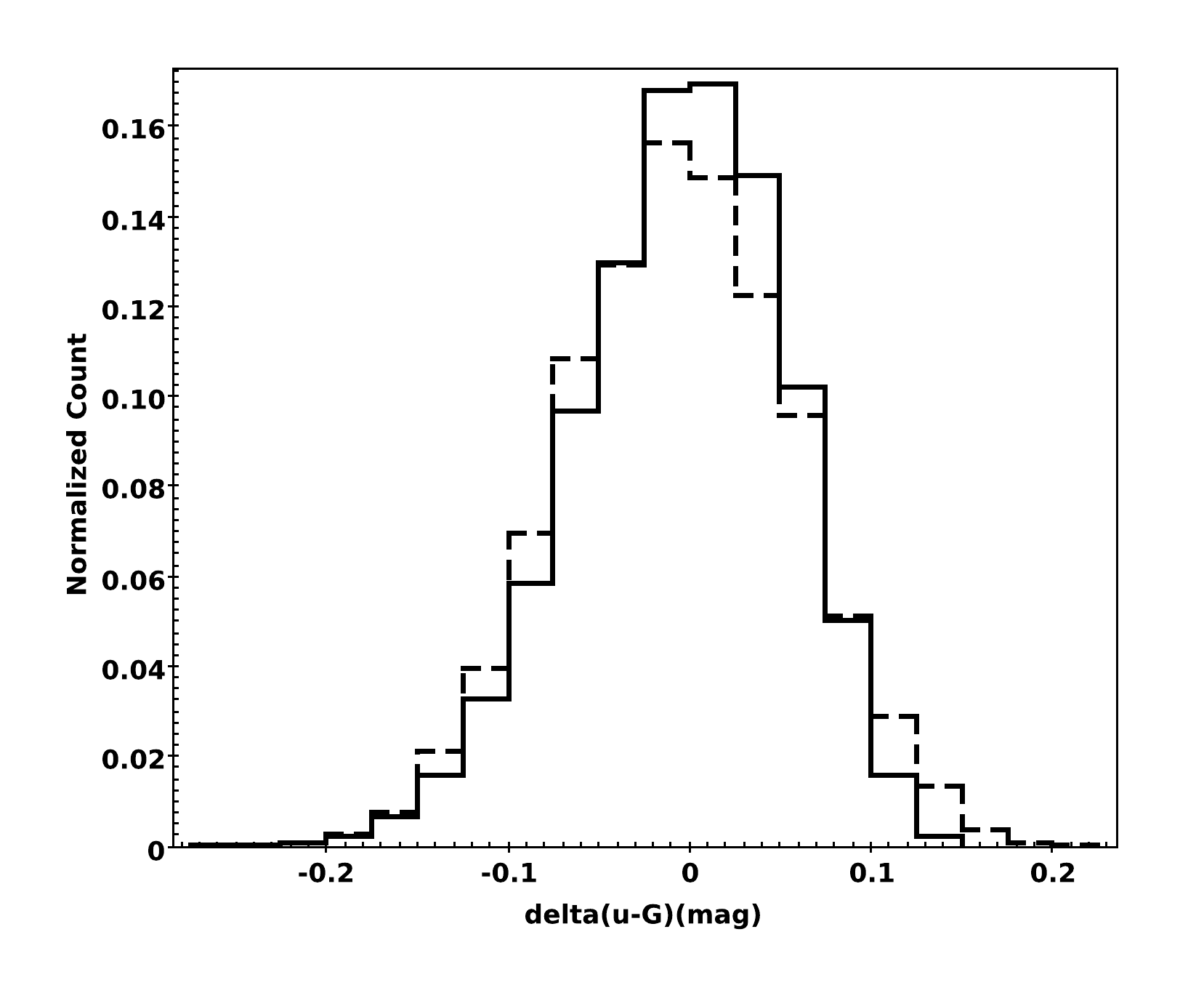}
\caption{ {Sensitivity of results to interpolation.  The histogram shows, for the $T_{eff}=4437~K$ star(left), and for the $T_{eff}=8472~K$ star (right) in
u-band, the effects of eliminating a randomly chosen third of the data points used for interpolation.  Solid histograms have the full interpolation grid, while the dashed have points removed. The u-band has been chosen because the $C_{uG}$ color is the most sensitive to stellar parameters.} 
\label{fig:InterpolationSensitivity} }
\end{figure}

Because our results are dependent on the interpolation procedure, and the grid of Phoenix spectra over which we interpolate, we performed an experiment to evaluate our sensitivity to this.  If the results are sensitive to the interpolation grid, eliminating a fraction of the grid should significantly change the color histogram.  The results of eliminating a random third of the grid is shown in Figure \ref{fig:InterpolationSensitivity}. As expected, the low temperature case shows greater sensitivity to interpolation errors than for the high temperature case, but both sensitivities are small in our context.  We conclude that the interpolation procedure does not significantly affect the reported results.

The achieved color uncertainties are much better at high temperature than low, particularly for the u and g bands, but all cases exceed the 5 mmag uncertainty requirement for LSST by factors of two (r-band) to ten (z-band).  It may be possible to improve these uncertainties by supplementing Gaia data with ground based spectroscopy.  An examination of Table 
\ref{table:Apsis_accuracy} shows that the accuracy is being degraded by low signal to noise for stars in our magnitude range. The higher resolution and signal-to-noise available from ground based instruments will largely eliminate this issue.  As an example of what could be obtained from the ground, \cite{Posbic2013} report accuracies for F and G dwarfs of 8K for $T_{eff}$, 0.08 for $\log g$ and 0.02 for $Z$, compared to the 160K, 0.3, and 0.3 respectively that we assumed from Apsis for similar stars.  This is certainly encouraging, but the high signal to noise and higher resolution required for such results would exact a price in telescope time, unless suitable spectra are already in the public domain.  The Gaia-ESO survey \citep{Smiljanic2014} may already have suitable data, though it is not yet publicly available. Error levels for FGK stars are 55K for $T_{eff}$, 0.13 for $\log g$ and 0.07 for $Z$, with possible systematic errors above these values. 

It is too soon to rule out the use of MS stars for LSST standards when they are supplemented with ground based spectroscopy, although there are still significant hurdles to clear.  Reddening will be a more significant issue for these stars than the WD, because of their much greater distances.  Additionally, we expect that the complexities of MS spectra may result in systematic errors in the transformation to the LSST bands which dwarf the errors resulting from stellar parameter determination.


\begin{table}
\begin{center}
\caption{Color uncertainties for main sequence stars.  Note that the assumption of a diagonal covariance matrix makes these uncertainties upper limits, with the true values likely lower by a factor of 2 to 4.  With the exception of r-G, the uncertainties still exceed the requirements by a significant factor.
\vspace{10pt}
\label{table:MS_SD}
}
\begin{tabular}{c|c|c}
Color & $\sigma$ (4437 K)(mag) & $\sigma$ (8472 K) (mag) \\
\hline
  u-G   &      0.269 & 0.057 \\
  g-G   &      0.099 & 0.037 \\
  r-G   &      0.010 & 0.014 \\
  i-G   &      0.037 & 0.040 \\
  z-G   &      0.061 & 0.048 \\
  y-G   &      0.074 & 0.045 \\
\end{tabular}
\end{center}
\end{table}

\FloatBarrier
\subsection{Using Gaia Data on WD Stars to Calibrate LSST}
\label{sec:UsingWD}
As has been emphasized by a number of authors (\cite{Holberg2007}, \cite{Holberg2006}), DA white dwarfs offer a unique opportunity for accurate photometric calibration of large surveys.  This is due to the fact that their intrinsic SED can be accurately calculated given only the stellar parameters $T_{eff}$ and $\log g$, both of which can be determined from ground-based spectroscopy of the Balmer line profiles.  This in turn enables the direct calculation of $C_{bG}$ from equation \ref{eqn:GaiaLsstColor}.

There are two issues that arise in putting this approach into practice for LSST.  First, DA WDs are relatively rare, since they cool rapidly through the temperature ranges where they are useful for photometric calibration.  Second, as discussed earlier, photometric standards for LSST must be relatively faint: $18 > r > 16$. DA WDs in this magnitude range can be significantly affected by interstellar extinction, which can be difficult to measure accurately from the ground.

The selection of DA WD to be used as photometric standards must account for three factors beyond the $18 > r > 16$ magnitude restriction.  WD with temperatures below about $13000K$ are likely to be variable and/or suffer from systematic errors in the determination of $\log g$ \citep{Eisenstein2006}. Those with temperatures above about $40000K$ are likely to have metals in the photosphere which make it impossible to utilize the pure hydrogen DA SEDs. Finally, the accuracy with which $T_{eff}$ and $\log g$ can be determined rapidly degrades as $T_{eff}$ increases.  We therefore choose the range $15000K < T_{eff} < 30000K$ for our candidates.

\begin{figure}
\plotone{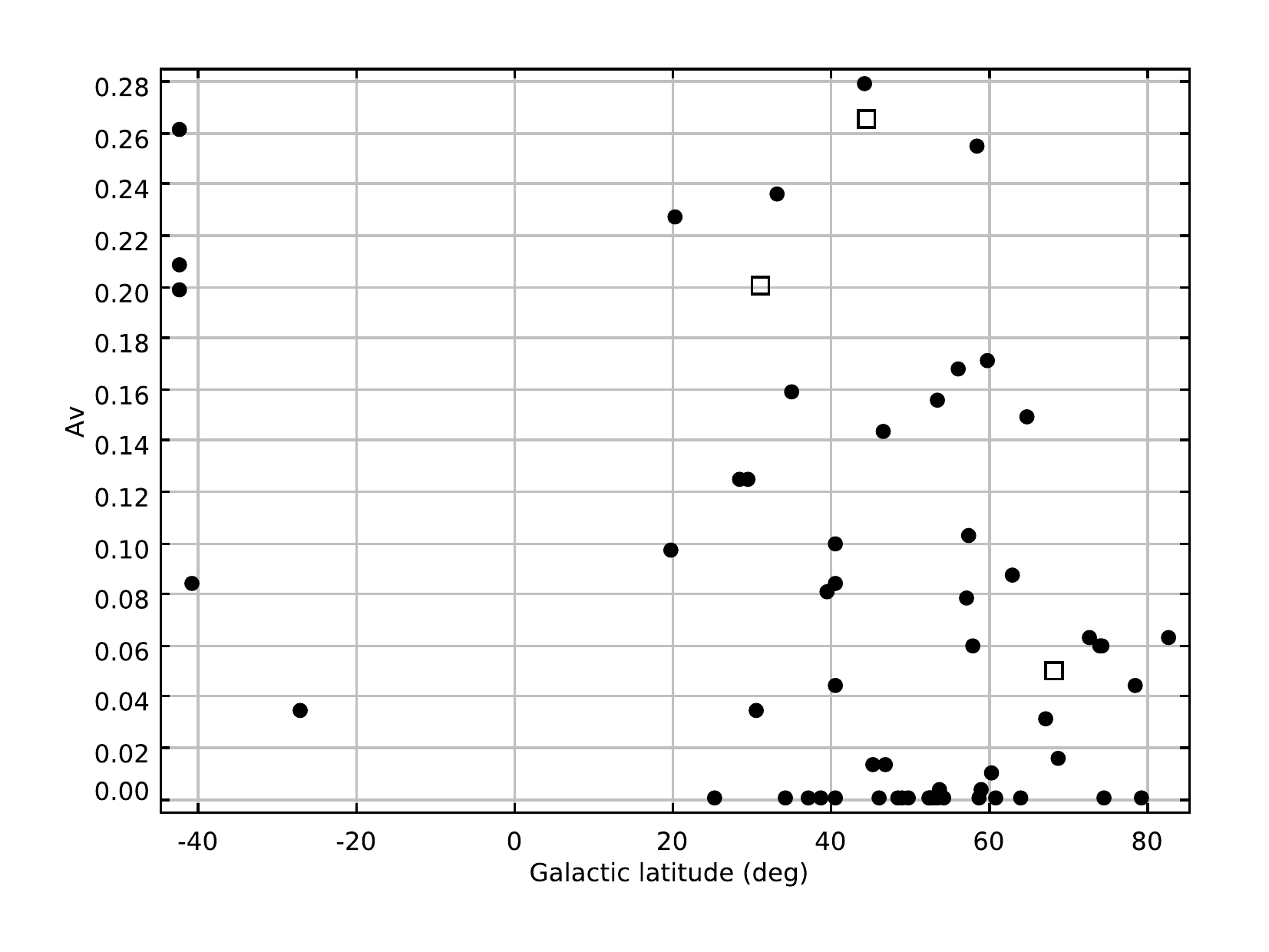}
\caption{Observed reddenings of faint WDs from \cite{AllendePrieto2009} (circles) and preliminary HST data from the program of \cite{Saha2013} (squares)}
\label{fig:WDReddening}
\end{figure}

To roughly estimate the number of WD which meet our selection criteria, we have used the Besancon model of the Galaxy \citep{Robin2003}.  Near the Galactic pole, the model predicts about $0.3 deg^{-2}$ within our magnitude range, and about $0.1 deg^{-2}$ additionally falling within our $T_{eff}$ range.  This is about 1 WD per LSST field, with larger numbers at lower Galactic latitude. The intrinsic faintness of WD means that our selected WD are relatively nearby, most within about 150 pc. As shown by Figure \ref{fig:WDReddening}, their reddening varies within the range $0<A_V<0.3$ mag, with stars at high Galactic latitudes ($b> 60 deg$) having generally $A_V<0.06$ mag. 

\subsection{Results for WD Stars}
The calculational procedure for determining the uncertainty of $C_{bG}$ for WD is nearly identical to that used for MS stars, as described in Section \ref{sec:MS_Results}.  The only differences are the methods for obtaining the stellar parameters and their uncertainties, and the fact that the stellar parameter space is of lower dimensionality because metallicity is no longer variable. 

There is now substantial information on the accuracy of stellar parameters derived from WD spectra.  We use values from \cite{AllendePrieto2009} and \cite{Barstow2003}.  We pick a fiducial DA WD with $T_{eff}=24000$, $\log g=8.4$, and $A_V=0.15$, with associated uncertainties $\sigma_T=100$, $\sigma_{logg}=0.01$.  We initially make the conservative assumption that $A_V$ is not determined by a stellar parameter pipeline, and instead just apply a uniform random distribution of $A_V$ in the range $0<A_V<0.3$ mag, consistent with the measurements in Figure \ref{fig:WDReddening}. We separately consider the case of high Galactic latitude WD, for which we assume, again from a uniform distribution, $0<A_V<0.08$ mag. The WD SEDs are generated from the TheoSSA web service \citep{Rauch2013}.

\begin{figure}
\plotone{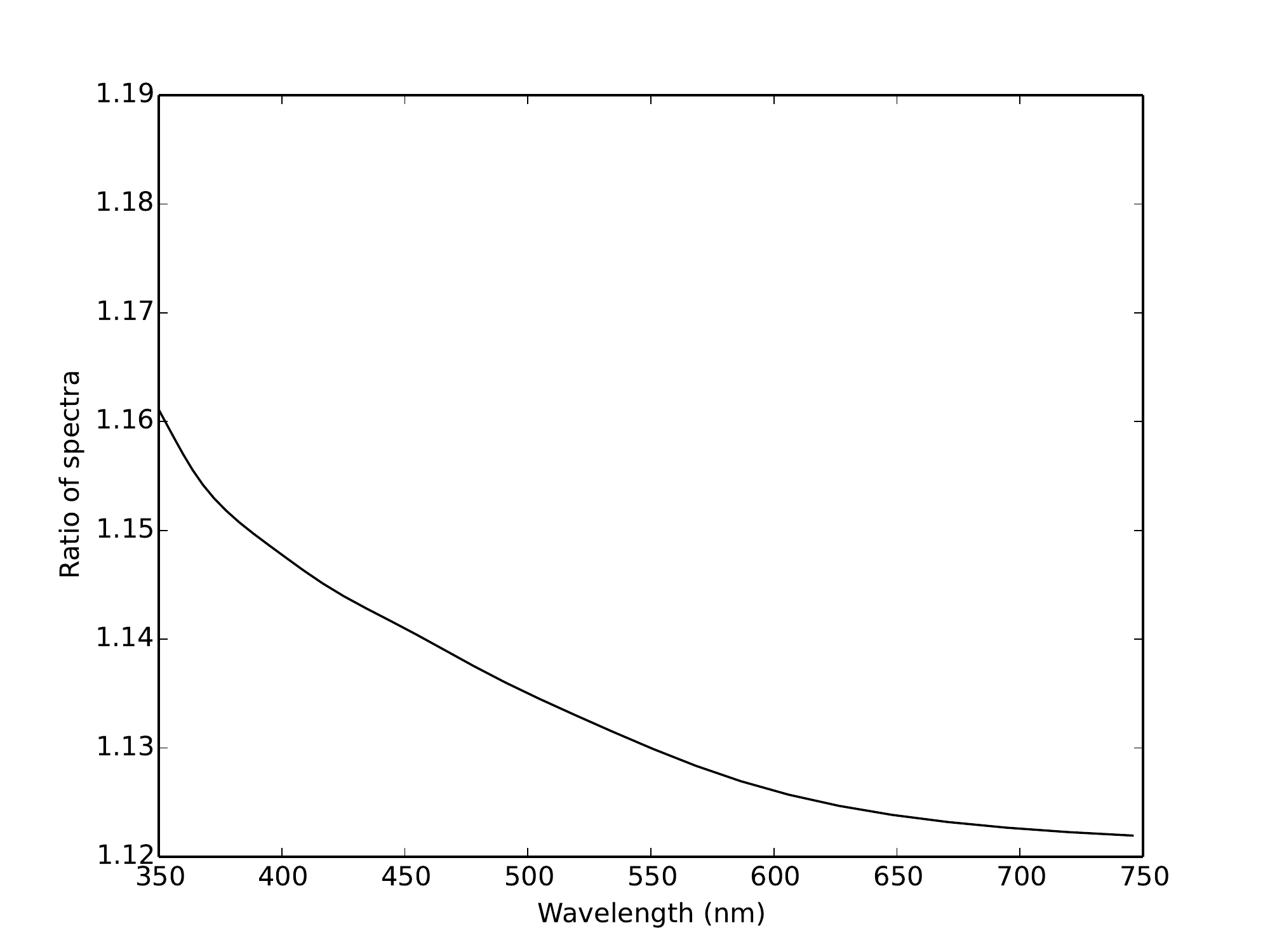}
\caption{ {The ratio of simulated Gaia BP spectra of a DA WD with $T_{eff}=20000$, $\log g=9$, and $A_V$=0 and 0.1. The spectral signature can plausibly be identified even in low signal to noise BP spectra.} 
\label{fig:BPreddening} }
\end{figure}

The results for our fiducial WD are summarized in Table 3.  For the WD without Galactic latitude restriction, the required accuracies for LSST standards are achieved in the r band, but exceeded in all other bands by factors of two to four.  The high latitude WD meet the requirements in all bands. There are at least two qualifications to be note with regard to these results.  
\begin{itemize}
\item As discussed in Section \ref{sec:Transforming}, the uncertainties presented here account \textbf{only} for the accuracy of the color transformation from the Gaia G-band to the LSST bands, assuming perfect models for WD spectra.  Systematic errors in the WD models themselves are not accounted for.  At present \citep{Holberg2007} our ability to assess these errors is limited by the accuracy in the available photometry.  HST observations underway, and of course Gaia itself, will allow these errors to be better quantified.
\item We have assumed lack of reddening information about individual objects beyond the overall range of the reddening, an assumption which is surely too pessimistic.  It is possible, for example, to determine the reddening from the Gaia RP and BP spectra once the WD SED has been determined from the ground based measurements of the Balmer lines.   Figure \ref{fig:BPreddening} shows the ratio of two simulated BP spectra of a WD with $T_{eff}=20000$, $\log g=9$, one with $A_V=0$ and one with $A_V~0.1$. The predicted per-pixel SNR of the BP spectra is about 40 at $G=18$ \citep{Bailer-Jones2013} and the reddening signature affects many pixels, which makes determination of the reddening to better than 0.1 seem plausible.  If this proves to be practical, the color uncertainties for the lower latitude WD would be correspondingly improved, likely bringing them within the requirements.
\end{itemize}

\begin{table}
\begin{center}
\caption{Color uncertainties for a DA WD with T=24000K, $\log g=8.4$.  $A_V$ is distributed uniformly between 0 and 0.3mag, or between 0 and 0.08 for $b>60$ deg.
\vspace{10pt}
}
\begin{tabular}{c|c|c}
Color & $\sigma$ (mag) & $\sigma$ ($|b| > 60\degr$)\\
\hline
  u-G & 0.027 & 0.008 \\
  g-G & 0.009 & 0.002 \\
  r-G & 0.004 & 0.001 \\
  i-G & 0.011 & 0.003 \\
  z-G & 0.015 & 0.004 \\
  y-G & 0.018 & 0.005 \\
\end{tabular}
\end{center}

\label{table:WD_SD}
\end{table}

\FloatBarrier
\section{Creating Synthetic Gaia G band Data from LSST Data}
\label{sec:SyntheticG}
\begin{figure}
\plotone{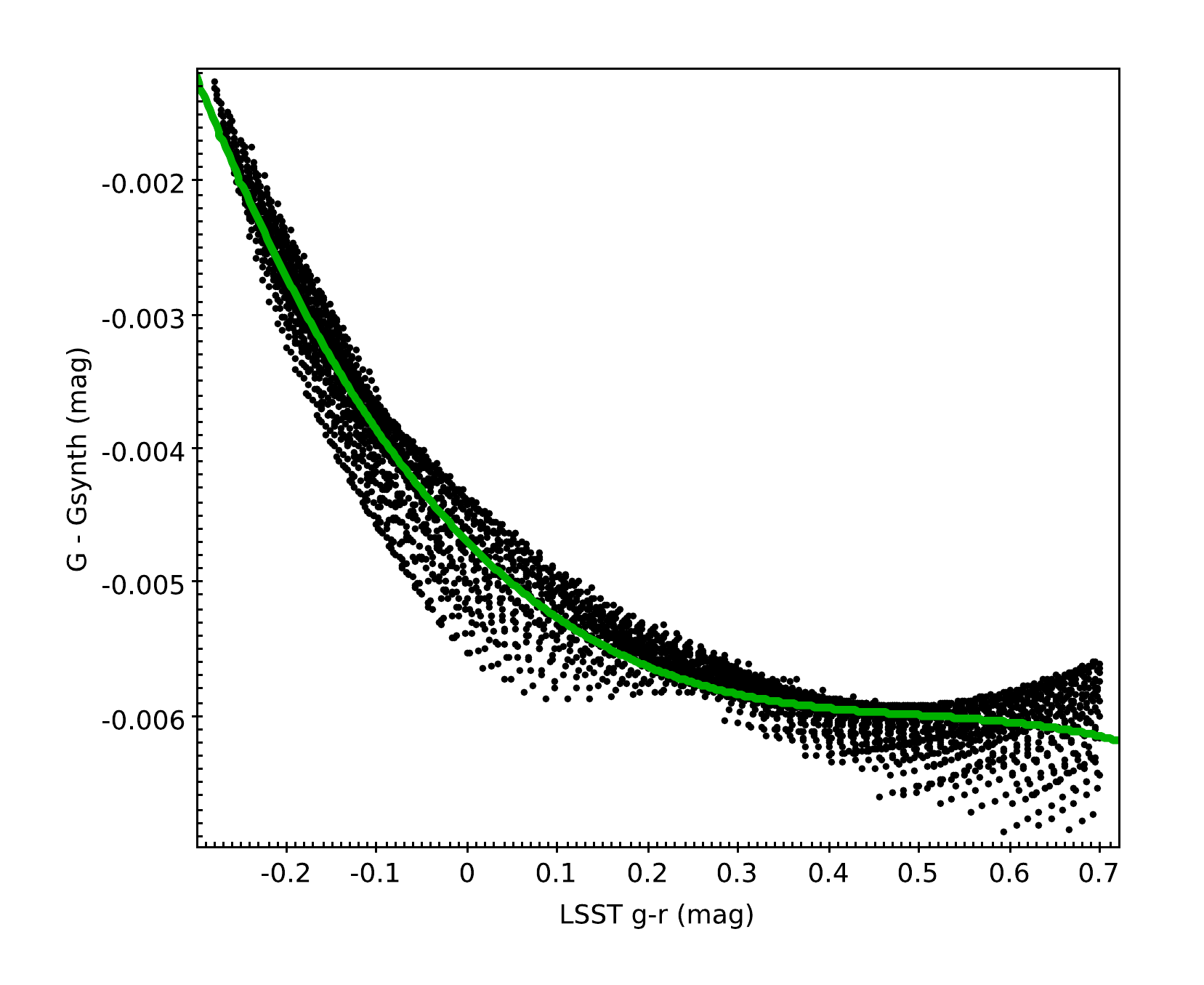}
\caption{ {Synthetic Gaia G residuals as a function of LSST g-r.  The solid curve is a least squares fit with a third order polynomial.} 
\label{fig:GaiaGSynthResid} }
\end{figure}

In this section we turn to the second approach mentioned in Section \ref{sec:LSSTPhotoCal}, that the six LSST bands can be combined into an approximate Gaia G band, and then compared directly with Gaia G band photometry, a possibility suggested by \cite{Ivezic2014}.  Looking again at Figure \ref{fig:bandpasses}, it seems plausible that the Gaia G bandpass can be reasonably approximated by linear combination of the LSST bandpasses
\begin{equation}
\phi_G(\lambda)~\approx~\sum_{i} \alpha_{i} \phi_{i}(\lambda)~\equiv~\phi_G^{synth}(\lambda)
\end{equation}
which leads to
\begin{equation}
F_G~\approx~\sum_{i} \alpha_{i} F_{i}~\equiv~F_G^{synth}
\end{equation}
where $F~=~\int f(\lambda) \phi(\lambda) d\lambda$ is the flux of an object with SED $f$ integrated over the bandpass $\phi$.  If the approximation were an equality, then we could directly compare, for all objects in the Gaia catalog within the LSST magnitude range, this synthetic G band flux to the actual measured G band flux.  Any discrepancies beyond the expected random scatter would indicate a systematic problem with the LSST photometric calibration process.  The spatial dependence of these discrepancies, and their dependence on stellar parameters, particularly $T_{eff}$, would give clues to what aspects of the calibration process were at fault.  Additionally, it might also be possible to correct the LSST fluxes to minimize the discrepancies, while maintaining consistency with our knowledge of stellar populations in color-color space, thereby improving the LSST calibration.
\noindent
In practice, there will be a significant difference between $\phi_G^{synth}$ and $\phi_G$, so that
\begin{equation}
\phi_G(\lambda)~=~\sum_{i} \alpha_{i} \phi_{i}(\lambda)~+~r(\lambda)
\end{equation}

leading to a systematic flux error
\begin{equation}
\delta F_G^{synth}~=~\int f(\lambda) r(\lambda) d\lambda
\end{equation}
Figure \ref{fig:GaiaGSynthResid} shows $\delta F_G^{synth}$ as a function of $g-r$, expressed in magnitudes, evaluated over the entire Phoenix SED set restricted to stars with $g-r~<0.7$.  This restriction corresponds roughly to $T_{eff}>5000 K$.  Cooler stars show much larger residuals.  As shown in the figure, if the $G_{synth}$ magnitudes are corrected with a third order polynomial in $g-r$, then the remaining scatter is $\sigma=0.2$ mmag, negligible in the LSST context.  Note that the same approach could be used for WD, with a smaller scatter, but also a much smaller sample of stars.

It is clear from this that $G_{synth}$ magnitude residuals will be diagnostic of any systematic errors in LSST calibration that exceed the 5 mmag level required by the SRD.  If a significant level of systematic error is found, the $G_{synth}$ data can be employed to diagnose the causes, though not as directly as with the direct standards discussed in Section \ref{sec:DirectGaiaStandards}.  

Could we use the $G_{synth}$ data to directly constrain, and thereby improve, the self calibration process, as we can with the direct standards? The main limitation is that, given an error $\delta F_G^{synth}$ for some particular star, there is not a unique way to decompose it into the errors in the individual LSST bands.  In the six dimensional LSST photometry space, we can move it anywhere on the five dimensional hyperplane given by
\begin{equation}
\sum_{i} \alpha_{i} \delta F_{i} = \delta F_G^{synth}
\end{equation}
To remove this degeneracy, we can additionally require that the star must remain on the color-color track for its stellar population.  To the extent that this track is scatter free, the corrections to the individual LSST magnitudes can be uniquely assigned.  Again, WD may be more valuable than MS stars for this purpose.  A quantitative assessment of the resulting accuracy is beyond the scope of this paper.

\begin{figure}
\plotone{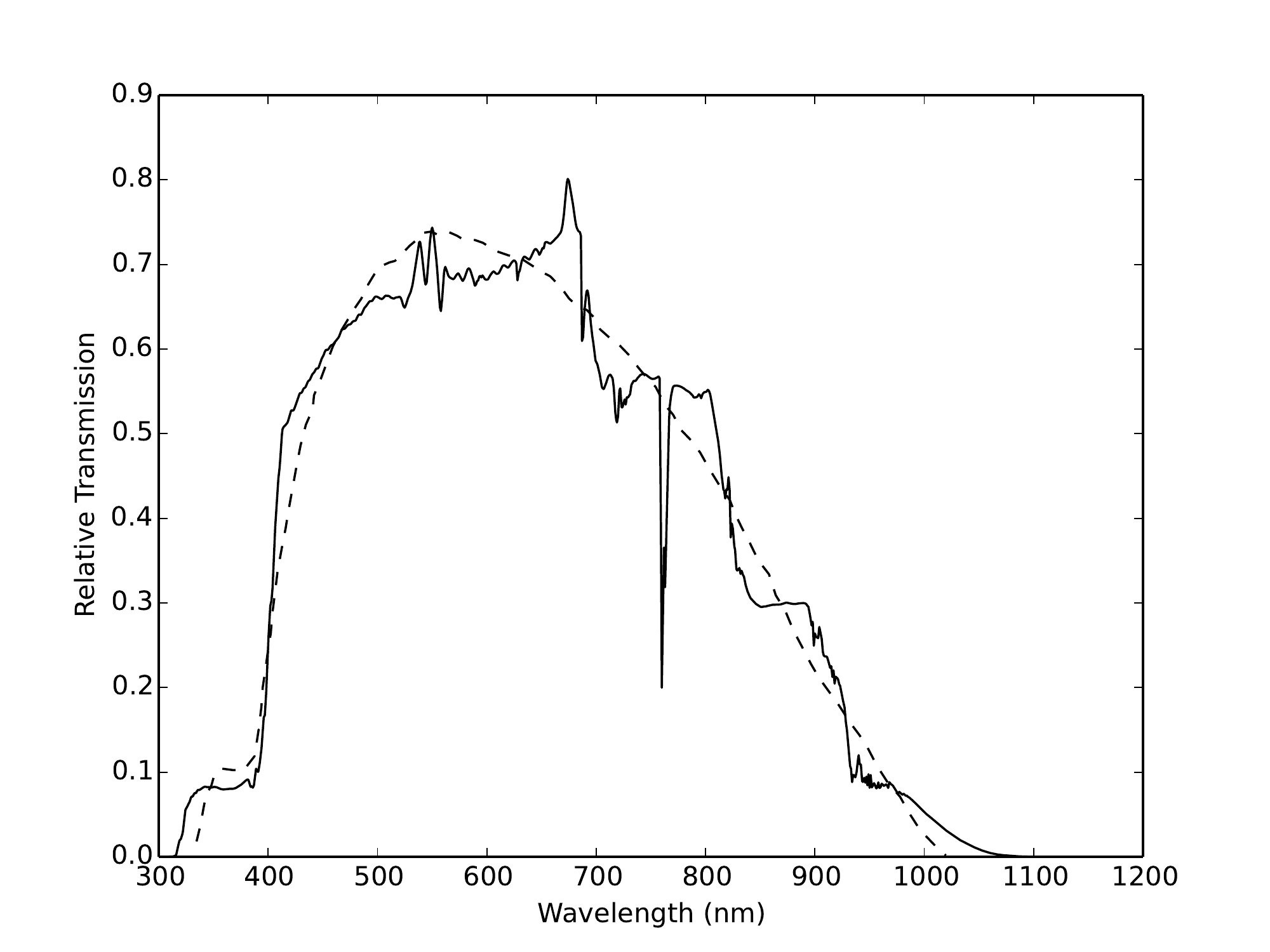}
\caption{ {Gaia G (dotted curve) and synthetic Gaia G (solid curve) bandpasses} 
\label{fig:GaiaGSynth} }
\end{figure}

\FloatBarrier
\section{Conclusion}
\label{sec:Conclusion}
We investigated two different approaches to using Gaia data to provide accurate photometric standards for wide-field surveys such as LSST.  In the first, Gaia G-band photometry is combined with stellar parameters derived from the Gaia Apsis pipeline (for MS stars) or from ground based spectroscopy (for WD) to determine the transformation from G-band to the six LSST bands.  The accuracy achieved with MS stars falls well short of LSST requirements, though with potential for improvement through the use of ground based spectroscopy.  Direct standards derived from WD have the potential to meet LSST requirements, particularly if they are restricted to high Galactic latitudes.  Additional work will be required to realize this potential in practice.  Possible systematic errors in WD SED models must be carefully assessed, and the accuracy of interstellar reddening derived from ground-based spectroscopy or from Gaia RP/BP spectroscopy needs to be more carefully quantified.  Both of these tasks will become possible in the near future using multi band HST photometry in conjunction with ground-based spectroscopy.
It is worth noting that the ground-based observations required to realize the potential of Gaia WD for photometric calibration may require substantial investment of telescope resources if a large number of standards is desired.  A spectrum of a single WD in our magnitude range requires roughly two hours of exposure time on a 6 meter class telescope to reach the necessary signal-to-noise.  Similar considerations apply to the use of MS stars.

In the second approach, the six LSST bands are linearly combined into a synthetic G-band which approximates the Gaia G-band.  Although the two G-bands have significant wavelength dependent differences, the resulting two sets of magnitudes have very small differences for most stars with $T_{eff} > 4000 K$.  This allows a useful check on the spatial dependence of LSST calibration accuracy, and within currently unknown limits, the correction of such systematic errors.

It seems worthwhile to actively pursue both approaches during the next years, as Gaia science data is released, and LSST nears operation.

\section{Acknowledgements}
We wish to thank Drs. Abhijit Saha (NOAO), Edward Olszewski (U. Arizona), and {\v Z}eljko Ivezi{\'c} (U. Washington) for many useful discussions which have improved the paper.

The TheoSSA service (http://dc.g-vo.org/theossa) used to retrieve theoretical spectra for this paper was constructed as part of the activities of the German Astrophysical Virtual Observatory.


\begin{figure}
\plotone{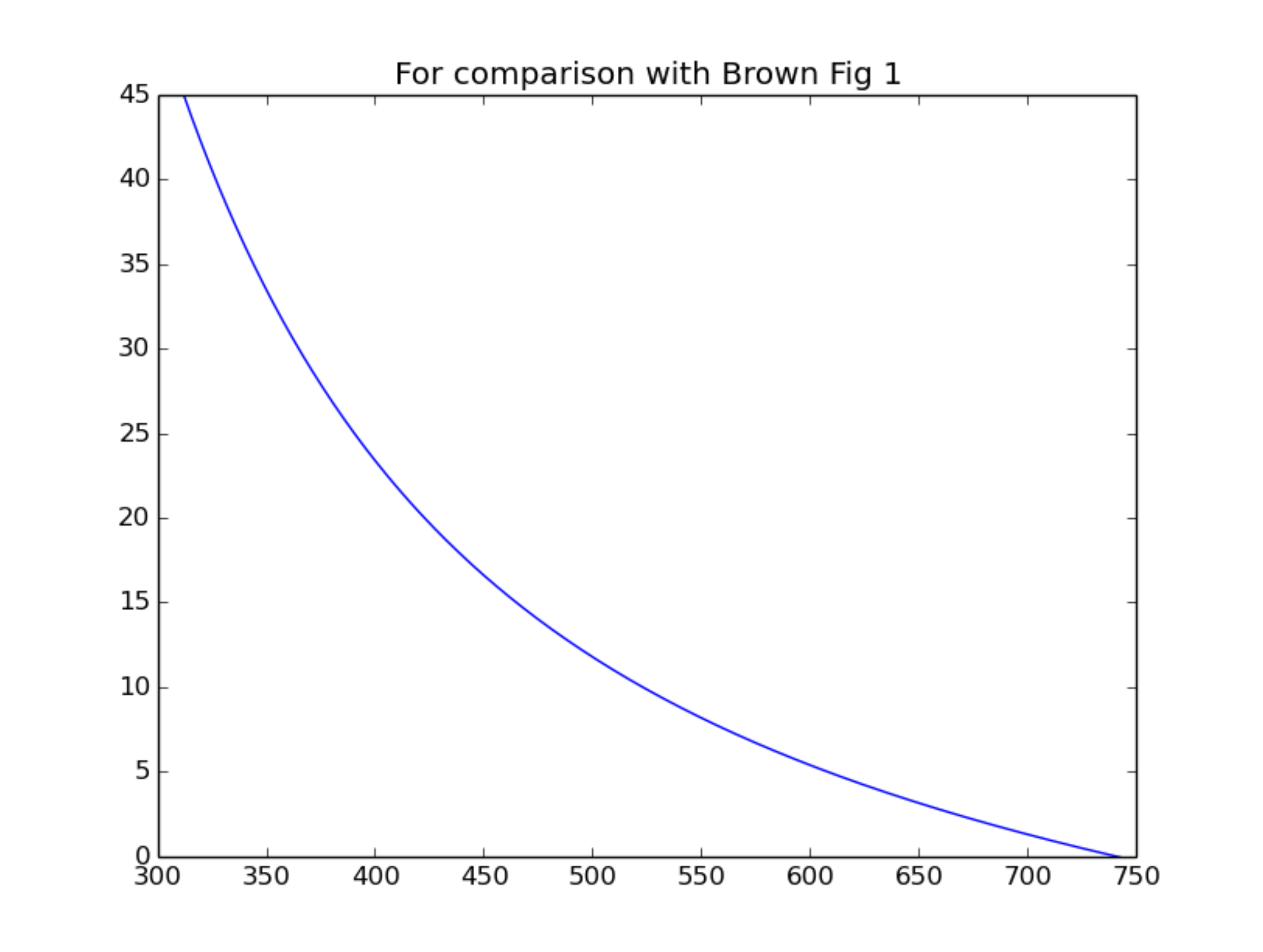}
\caption{Dispersion function from the BP simulator.  Compare with Brown 2006, Fig 1.  }
\label{fig:dispersion}
\end{figure}
\begin{figure}
\plotone{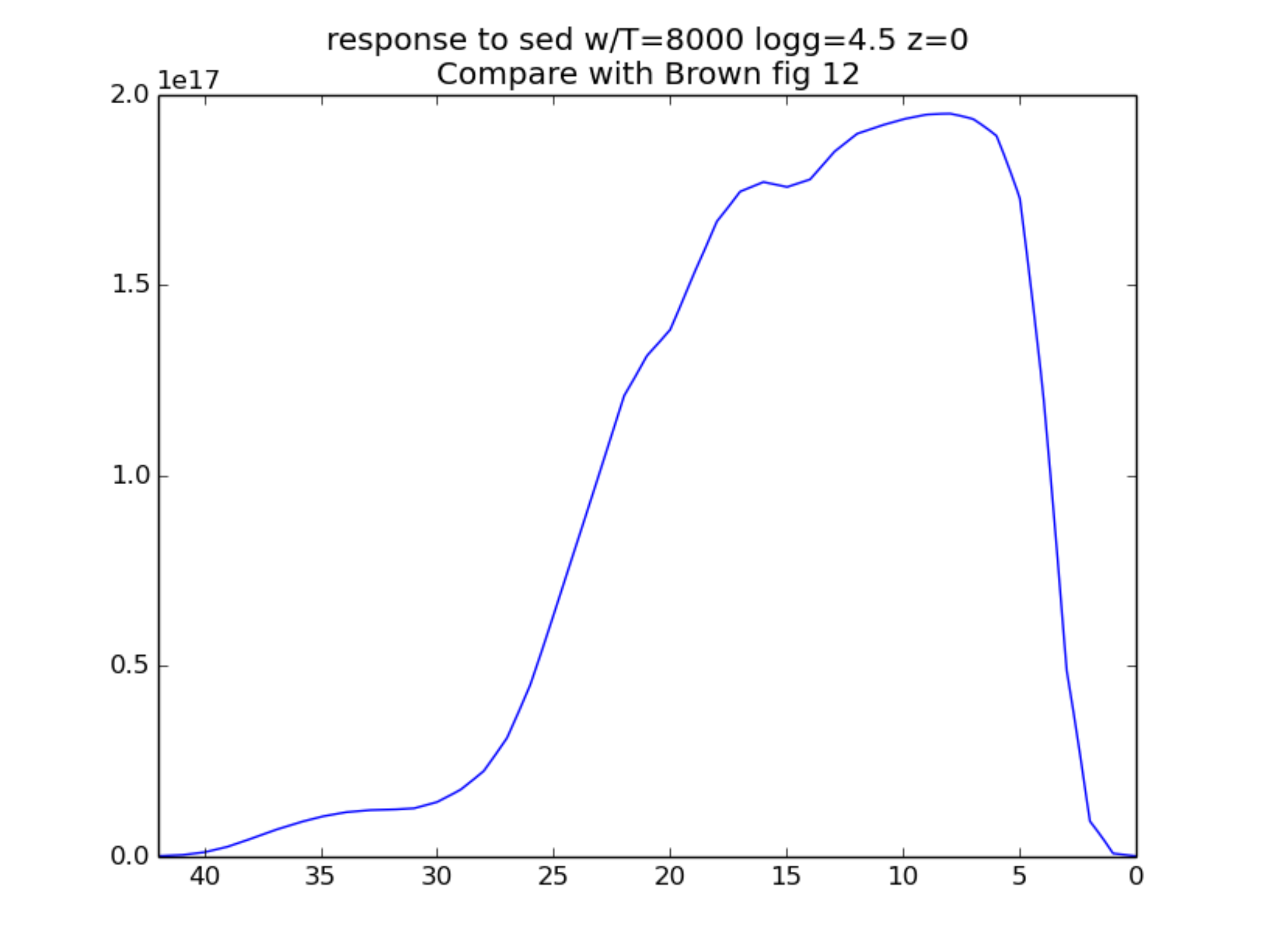}
\caption{Simulated BP response to a flat SED.  Compare with Brown 2006, Fig 5 }
\label{fig:flatSED}
\end{figure}
\begin{figure}
\plotone{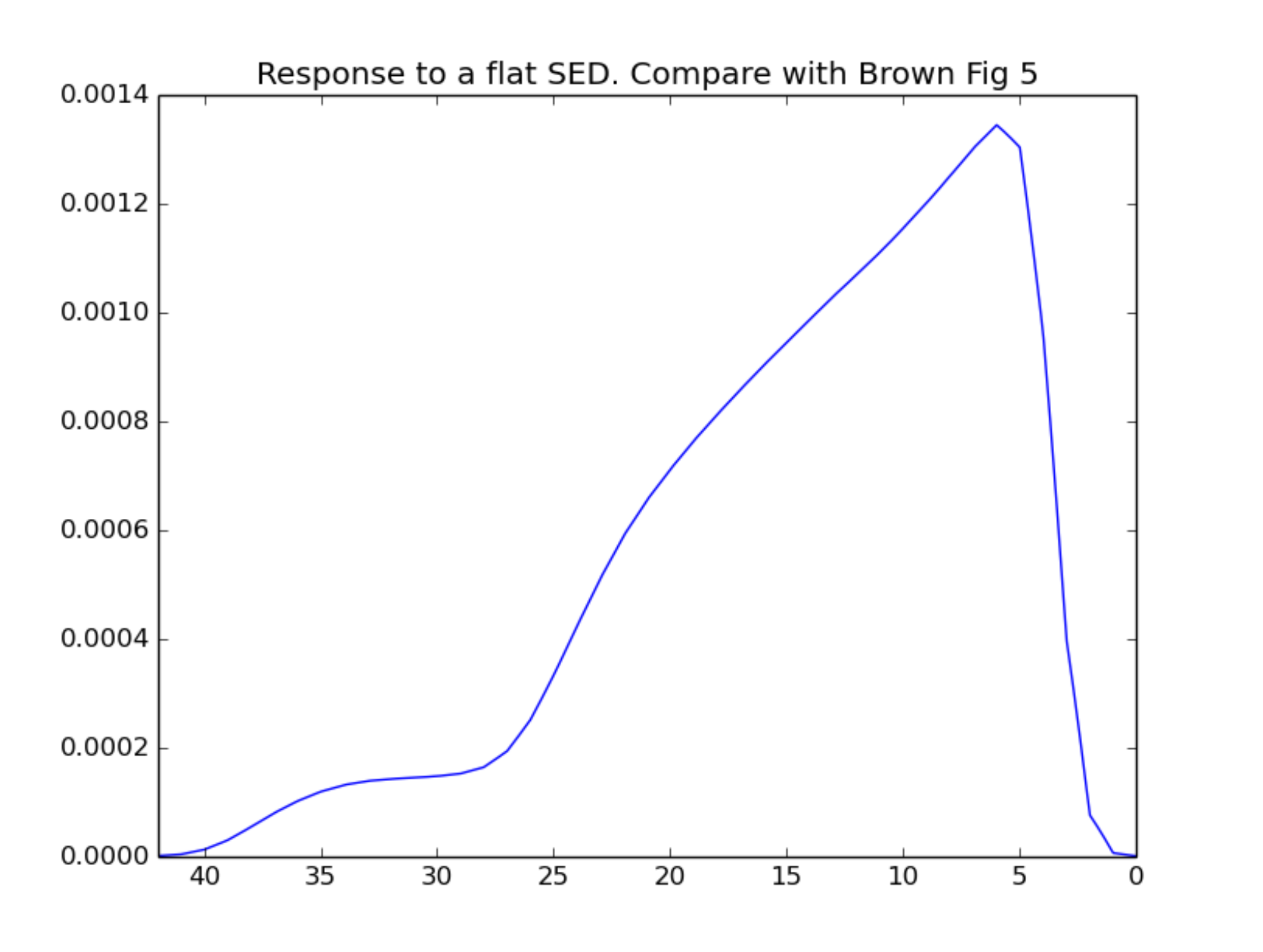}
\caption{Simulated BP response to a star with T=8000, logg=4.5 [Fe/H]=0. Compare with Brown Fig 12  }
\label{fig:dispersion}
\end{figure}
\end{document}